\documentclass[a4paper,12pt]{article}
\newcommand{\be}{\begin{equation}}
\newcommand{\ee}{\end{equation}}
\newcommand{\ben}{\begin{eqnarray}}
\newcommand{\een}{\end{eqnarray}}
\newcommand{\ba}{\begin{array}}
\newcommand{\ea}{\end{array}}
\newcommand{\al}{\alpha}
\newcommand{\cF}{{\cal F}}
\newcommand{\ep}{\epsilon}
\newcommand{\mat}[4]{ {\left( \ba{cc} #1 & #2 \\ #3 & #4 \ea \right)}}
\newcommand{\non}{\nonumber}
\newcommand{\del}{\partial}
\newcommand{\Gh}[2]{ {\Gamma_{\hat{#1}\hat{#2}} }}
\newcommand{\journal}[4]{{\rm #1} {\bf #2} (19#3) #4}
\newcommand{\NP}{\journal{Nucl. Phys.}}
\newcommand{\PL}{\journal{Phys. Lett.}}

\newcommand{\PR}{\journal{Phys. Rev.}}

\newcommand{\JHEP}{\journal{JHEP}}

\newcommand{\ATMP}{\journal{Adv.Theor.Math.Phys.}}
\makeatletter
 
 \@addtoreset{equation}{section}
\makeatother
\begin{document}

\begin{titlepage}
\begin{flushright}
       {\normalsize  OU-HET-340 \\
       YITP-99-79 \\ hep-th/9912266  \\
           December 1999}
\end{flushright}
\vspace{10mm}
\begin{center}
  {\Large \bf Deformation of Conifold \\ and \\ Intersecting Branes}

\renewcommand{\thefootnote}{\fnsymbol{footnote}}
\vfill
         { Kazutoshi ~Ohta}\footnote{e-mail address: kohta@yukawa.kyoto-u.ac.jp} \\
\vspace{2.5mm}
{\em Yukawa Institute for Theoretical Physics, Kyoto University,\\
Kyoto 606-8502, Japan}\\
\vspace{0.8mm}
            and \\
\vspace{0.8mm}
         { Takashi ~Yokono}\footnote{e-mail address: yokono@funpth.phys.sci.osaka-u.ac.jp}\\
         \vspace{2.5mm}
   {\em     Department of Physics,
        Graduate School of Science, Osaka University,\\
        Toyonaka, Osaka, 560-0043 Japan}\\
\renewcommand{\thefootnote}{\arabic{footnote}}
\end{center}
\vfill
\begin{abstract}
We study the relation between intersecting NS5-branes whose intersection is 
smoothed out and the deformed conifold
 in terms of the supergravity solution. 
We  solve the condition of preserved supersymmetry on a metric inspired by the deformed conifold metric and obtain a solution of the NS5-branes  which is delocalized except for one of the overall transverse directions.  
The solution has consistent properties with other configurations obtained by
 string  dualities.
\end{abstract}
\vfill
\end{titlepage}

\newpage

\section{Introduction}

A system of parallel D3-branes at a conifold \cite{AK, KW, AFHS, MP} has been discussed from the viewpoint of AdS/CFT 
correspondence \cite{maldacena} (and for a review \cite{AGMOO}). 
From the properties of the conifold, we can identify the field theory on the 
N D3-branes with ${\cal N}=1$ gauge group $SU(N)\times SU(N)$ theory with a quartic superpotential 
in the infrared \cite{KW}.

By using T-dualities, the system is mapped to  
configurations of D-branes and 
 intersecting NS5-branes over a 1+3 dimensional 
world-volume \cite{DM,Uranga}. 
In the brane picture, one can intuitively read out the gauge group and the 
spectrum of the microscopic theory on the D-branes.

The T-duality relations between intersecting NS5-branes 
and the conifold play an
important role in order to map the system of D3-branes at the conifold singularity to 
the brane configurations.

The  relation between the NS5-branes and the conifold  is discussed in 
terms of supergravity solutions by performing  the T-duality along one of the overall directions of 
the  NS5-branes \cite{DM}.
The T-duality maps the field 
theory on  the D3-branes at the 
conifold singularity onto D4-branes which are stretched between both sides of NS5-branes
 and extend along the compactified direction.
 Such configurations are called  elliptic models \cite{witten}. 
 The duality relation is generalized to other  conifold types and various 
 types of NS5-branes \cite{Uranga, Unge, AKLM, TO}.

There is another T-duality
relation between the intersecting NS5-branes 
and the conifold \cite{BSV}.  
The NS5-branes are mapped to the conifold by performing two T-dualities along 
one of the  relative transverse directions of each NS5-brane. 
By taking these T-dualities, 
one has D5-branes which fill the compact brane box \cite{HZ, HU}. 
It is discussed that the intersecting point of NS5-branes must be resolved to 
obtain the field theory on the D5-branes which has suitable 
gauge group and superpotential \cite{AKLM}.  The intersecting NS5-branes 
are also described by 
 a single NS5-brane wrapping the curve which supports a non trivial $S^1$
  that D5-branes can end. 
Such NS5-brane is called NS5-brane with a ``diamond''. 
In the conifold picture, the conifold singularity 
  is resolved.

In the absence of D-branes,  local mirror symmetry is proposed between 
 generalized  and orbifolded conifolds \cite{AKLM}.  
 The mirror transformation is equivalent to the T-duality on the 
 supersymmetric toroidal 3-cycle which Calabi-Yau manifolds and their mirror manifolds equip 
 \cite{STZ}. In the conifolds picture, the T-duality is 
 equivalent to  the combination of 
 above two types of T-dualities.  Performing the mirror transformation to the 
 blownup conifold, 
 one find that the deformed conifold is mapped to the NS5-brane with the diamond by the T-duality along one of the overall transverse  directions.

Our purpose in the present paper is to explore the relation between the 
NS5-branes with the diamond and the deformed conifold in  terms of 
supergravity 
solutions. 
The guideline to obtain the deformed conifold metric 
is discussed in \cite{CO} and the explicit metric is presented  
in \cite{MT}.  We start with a metric inspired by the deformed 
conifold metric. Solving the condition of the preserved supersymmetry, after
 some replacements of line elements \cite{DM} and the T-duality,  we will
  have a solution of the NS5-branes 
with a diamond which is 
delocalized except for one of the overall transverse directions. 
As a result, we find that the size of the diamond relates to the displacement of end points of a D4-brane on  a NS5-brane. We confirm it by using string 
dualities.

This paper is organized as follows. In sec. \ref{secpre}, we present  a short 
summary on the duality relations between metics of the NS5-branes and the conifold  \cite{DM}, and discuss the deformation of the conifold algebraic geometry 
 \cite{AKLM}. We also explicitly give the deformed conifold metric \cite{CO, MT}. 
In sec. \ref{secmetric}, we  construct  the NS5-branes with the diamond metric 
and  
show that the metric is also obtained from the intersecting NS5-branes metric 
by some coordinate transformation.  
In sec. \ref{secdual}, we make clear the meaning of  the size of the diamond by using
 some string dualities.  
 Sec. \ref{secconclusion} is devoted to conclusion and discussion.

\section{NS5-branes and Conifold \label{secpre} }
\subsection{Duality relations}

Let us start with  intersecting NS5 and NS5'-branes whose world-volume directions are
\be
\ba{c|cccccccccc}
NS5 &( 0 & & & & & 5 & 6 & 7 & 8 & 9 )\\
NS5'&( 0 & & &3&4&  &    &7 & 8 & 9 ).
\ea
\ee
We consider the intersecting NS5-branes metric  smeared except for the $x^1$-direction. The metric relates to the conifold metric by the T-duality along the $x^2$-direction \cite{DM}. 

 The conifold is topologically a cone over a 5-dimensional base manifold $S^2 \times S^3$. To see the relation between NS5-branes and the conifold, it is 
 useful to consider the geometry of the base manifold as 
 $U(1)$ fibration over a base $S^2 \times S^2$. 
 In the intersecting NS5-branes background, we have a similar geometry which is $S^1$  over 
 $R^2 \times R^2$. Here $S^1$ is the $x^2$-direction which is 
 compactified to take the T-duality, and $R^2 \times R^2 $ is the $(x^3,x^4)$ and $(x^5,x^6)$ directions which are planar. Therefore we must  ``compactify'' the directions in order to fit the topology to the conifold. This is done by replacing the Mauer-Cartan 1-form of $R^2 \times R^2 $ by of $S^2 \times S^2$, that is,
\ben
dx^{3,5} &\to & \sin\theta_{1,2}\, d\phi_{1,2} ,\\
dx^{4,6} &\to & d\theta_{1,2}.
\een

After the T-duality,  we have the conifold metric below up to coefficients with certain replacements, for example $x^2\,  \to\, \psi$ where $\psi $ is the coordinate on the $U(1)$ fiber.
It is difficult to determine the coefficients from the intersecting NS5-branes metric  because the metric is delocalized except for only one direction though  the conifold metric localizes in the all directions. 

The conifold metric is given in \cite{CO},
\be
ds^2_{conifold} = dr^2 + r^2\, \left( \frac{1}{6}\sum_{i=1}^2 (d\theta_i + \sin^2\theta_i\, d\phi_i^2 ) + \frac{1}{9}(d\psi + \cos\theta_1 d\phi_1 + \cos\theta_2 d\phi_2)^2 \right). \label{conifold}
\ee
Here $r$ is a radial coordinate in the conifold and relates to the 
$x^1$-direction in the intersecting NS5-branes picture.

The relation between the generalized and orbifolded conifold whose singularities are resolved is discussed in \cite{AKLM}. The simplest example of it is the 
T-duality or mirror symmetry between   the deformed conifold and the 
blowup of conifold. 
The conifold has  vanishing  2-cycle and 3-cycle at the origin where is an 
isolated singular point. In the intersecting NS5-branes picture, 
the singularity
corresponds to a singular point where the NS5-brane intersects with the 
NS5'-brane. On the conifold  the singularity can be resolved by 
deforming the conifold.   In the NS5-brane picture, the deformation 
smooth out the intersection  with a non-vanishing cycle.

The conifold is algebraically defined by $x\, y = u\, v $, where $(x,y,u,v) \in {\bf C}^4$. The deformed conifold is described by
\be
xy = uv + \epsilon^2 \label{defconicurve}
\ee
where the singularity is resolved by a 3-cycle with non-zero radius $\epsilon $. 
By using the T-duality the deformed conifold maps to NS5-branes wrapping a 
curve which appears on $uv=0$  \cite{KLMVW},
\be
xy = \epsilon^2. \label{dia}
\ee
In the conifold case, $\epsilon $ vanishes and  the curve becomes $xy=0$. The solution   separates into $ x=0$ and $y=0$. Each equation describes a location of the NS5-brane. Hence 
the conifold simply maps to intersecting NS5 and NS5'-branes after the T-duality. This fact agrees with the discussion using the metric.  When the $\epsilon \ne 0$, the curve (\ref{dia}) is smooth and describes topologically a 
sphere  of radius $\epsilon $. We call the non-vanishing cycle which NS5-branes wrap as a ``diamond'' \cite{AKLM}.

\subsection{Metric of the deformed conifold\label{secdefconi}}
The metric of (\ref{defconicurve})  is determined from the condition that the metric is Ricci flat and K\"ahler \cite{CO, MT}, 
\ben
ds^2 &=& {\cal F}' tr(dW^\dag dW) + {\cal F}'' |tr(W^\dag dW)|^2. \\
\een
Here $W$ is a complex $2\times 2$ matrix which satisfies the condition 
corresponding to (\ref{defconicurve}),
\be
\det W = -\frac{1}{2}\epsilon^2.
\ee
Here  we fix $\epsilon $ as  a real parameter. 
We define a radial coordinate $\rho^2 $ in  ${\bf C}^4$ space as
\be
\rho^2 \equiv tr(W^\dag W).
\ee
${\cal F}={\cal F}(\rho^2)$ is a K\"ahler potential and ${\cal F}' \equiv \frac{d{\cal F}}{d(\rho^2)}$  is determined by the condition that the metric is Ricci flat as 
\be
\cF' = \ep^{-\frac{2}{3}}\, K .
\ee
Here $K$ is a function defined as
\ben
K(\tau ) &\equiv  & \frac{(\sinh 2\tau  - 2\tau )^\frac{1}{3}}{2^{\frac{1}{3}}\sinh\tau },\\
\rho^2 &\equiv & \epsilon^2 \cosh\tau \label{rtau}.
\een

We now take $W$ as
\ben
W &=& L\, W_{\epsilon}\, R^\dag, \\
W_{\epsilon } &=& \mat{0}{\frac{\sqrt{\rho^2+ \epsilon^2}+\sqrt{\rho^2-\epsilon^2}}{2}}{\frac{\sqrt{\rho^2+ \epsilon^2}-\sqrt{\rho^2-\epsilon^2}}{2}}{0}.
\een
The  $SU(2)$ matrix $L,R$ are parametrized in terms of Euler angles,
\be
\mat{\cos\frac{\theta_k}{2}\,e^{i\, (\psi_k + \phi_k)/2}}
{-\sin\frac{\theta_k}{2}\,e^{-i\, (\psi_k - \phi_k)/2}}
{\sin\frac{\theta_k}{2}\,e^{i\, (\psi_k - \phi_k)/2}}
{\cos\frac{\theta_k}{2}\,e^{-i\, (\psi_k + \phi_k)/2}}
\ee
where $k=1,2$ for $L,R$ respectively. 
The stability group of $W_\epsilon $ is a $U(1)$ which fixes 
$\psi_1 + \psi_2 \to \psi$. 

Eventually we have the deformed conifold metric,
\be
ds^2= K \ep^{\frac{4}{3}} \left( \frac{\sinh^3\tau}{3\, (\sinh 2\tau  -2\tau )} (d\tau^2 + ds_1^2)
+ \frac{\cosh\tau }{4} ds_2^2 + \frac{1}{4} ds_3^2 \right), \label{defconimet}
\ee
where 
\ben
ds_1^2 &\equiv & (d\psi + \cos\theta_1\; d\phi_1
 + \cos\theta_2\; d\phi_2)^2, \label{conids1} \\
ds_2^2 &\equiv & d\theta_1^2 + d\theta_2^2 + \sin^2\theta_1\; d\phi_1^2 + 
 \sin^2 \theta_2 \; d\phi_2^2, \label{conids2}\\
ds_3^2 &\equiv & 2\;\left( \sin \psi\,
          \left( {d\phi_1}\,{d\theta_2}\,
             \sin {\theta_1} + 
            {d\phi_2}\,{d\theta_1}\,
             \sin {\theta_2} \right) \right. \non \\
  && \qquad + \left. 
         \cos \psi\,\left( {d\theta_1}\,
             {d\theta_2} - 
            {d\phi_1}\,{d\phi_2}\,
             \sin {\theta_1}\,\sin {\theta_2}
             \right)  \right) \label{conids3}.
\een
The determinant of the deformed conifold metric is proportional to 
$\sinh^4\tau$  which vanishes if $\tau \to 0$, and the deformed conifold 
reduces to a lower-dimensional subspace.  From eq.(\ref{rtau}), 
the limit means that $\rho \to \epsilon$ where the stability group 
enhances to $SU(2)$. At $\rho =\epsilon $, the geometry becomes 
$(SU(2)\times SU(2))/SU(2) = S^3$. 
So the deformed conifold metric reduces to the $S^3$ surface metric.

If we take another limit,
\be
\rho^2 = \epsilon^2 \, \cosh \tau \quad \mbox{fixed as}\quad \tau,\, 1/\epsilon \to \infty,
\ee
the deformed conifold metric (\ref{defconimet}) reduces to the conifold one (\ref{conifold}) with a coordinate transformation $\rho^2 = (\frac{2}{3})^{\frac{3}{2}} r^3$.  Thus we confirm that when the size of 3-cycle $\epsilon $ vanishes the metric smoothly deforms to the conifold one.

The deformed conifold metric has the term (\ref{conids3}) which does not exist 
in the conifold case (\ref{conifold}). Since the deformed conifold is 
considered as the 
T-dual  of the NS5-branes with a diamond,  we expect that the additional term 
is closely related to an effect of the existence of diamond.

\section{NS5-branes with a diamond metric \label{secmetric} }

In this section, we consider the relation between the deformed 
conifold metric and the NS5-branes with  the diamond metric which  is 
smeared except for  one of the overall transverse directions $x^1$ as in 
the previous section. The metric of the NS5-branes with the diamond  relates 
to the deformed conifold metric after the T-duality.  In the latter case, 
there is no gauge fields and dilaton background in the corresponding string theory.  So  we focus only on the metric of the T-dual of 
NS5-branes for a while and solve the condition for a 
preserved supersymmetry on 
the metric.

The deformed conifold metric (\ref{defconimet}) is  fully localized.  On the other hand, however, we are looking for a smeared metric. 
In this case, we  assume that some replacements of the line elements
 would be again similar to the intersecting NS5-branes and conifold case,
\ben
\sin\theta_{1,2} d\phi_{1,2} &\to &dx_{3,5}, \label{replace1}\\
d\theta_{1,2} & \to & dx_{4,6},\label{replace2}
\een
and 
\be
d\psi \to dx_2.\label{replace3}
\ee

We now would like to take the T-duality along the $x^2$-direction. 
Hence we need a $U(1)$ isometry along the direction. However, 
we have functions depending on the $x^2$-coordinate which are $\sin x^2$ 
and $\cos x^2$ in the deformed conifold metric (\ref{defconimet}) in 
 the above replacements. 
We assume that they become some constants $a_1$ and $a_2$ after delocalization. 
So eqs.(\ref{conids1}), (\ref{conids2}) and (\ref{conids3}) become
\ben
ds_1^2 &=& (dx^2 + B_1 dx^3 + B_2 dx^5)^2 ,\\
ds_2^2 &=& \sum_{i=3}^6 (dx^i)^2 ,\\
ds_3^2 &=& 2\, ( a_1\, (dx^3\, dx^6 + dx^5\, dx^4 ) 
 + a_2\, (dx^4\, dx^6 - dx^3 \, dx^5 ) ). \label{ds3x}
\een
Here $B_1, B_2 $ are some functions which could not be determined
 by the above replacements even in the conifold case. They should be
   determined by solving the supersymmetry condition. 
We will specify the  functions by an  ansatz as we will see below.

Thus we consider the superstring compactification on the six-dimensional 
curved space which described by the following metric, 
\ben
ds^2 &=& g_{\mu\nu} dx^\mu dx^\nu \non \\
&=& A(x^1)^2\, (dx^1)^2 + B(x^1)^2\, ds_1^2 + 2 C(x^1)\, ds_2^2 + 2 D(x^1)\, ds_3^2 \, . \label{startmetric}
\een
Here $A(x^1), B(x^1), C(x^1)$ and $D(x^1)$ are functions depending only on  the $x^1$-coordinate. 
We have introduced $a_1$ and $a_2$ as arbitrary constants. If we rescale 
$\sqrt{a_1^2 + a_2^2}D(x^1)$ as $ D(x^1)$ and $ds_3^2/\sqrt{a_1^2 + a_2^2}$ as $ds_3^2$, the constants $a_1$ and $a_2$ become  $a_1/\sqrt{a_1^2 + a_2^2}$ and 
 $a_2/\sqrt{a_1^2 + a_2^2}$ in $ds_3^2$ (\ref{ds3x}). Therefore the redefinition means that  we simply set 
\ben
a_1 &\to & \sin\al ,\\
a_2 &\to & \cos\al
\een
where $\al $ is a constant.

Since we expect that above metric (\ref{startmetric})  becomes the solution of NS5-branes with a diamond after the T-duality along the $x^2$-direction, we make the following  ansatz before solving the  preserved 
supersymmetry condition:\\
(i)  The both $x^1$ and $x^2$-directions are the overall transverse directions 
to the NS5-branes after the  T-duality. Therefore the coefficients of $(dx^1)^2 $ and $(dx^2)^2$ must be the same after that. Since the metric $g_{\mu\nu}$   and the T-dualized metric 
$j_{\mu\nu}$  are related each other by
 $j_{11} = g_{11} $ and $j_{22} = 1/g_{22} $ \cite{BHO}, it  means that
\be
B(x^1) = A(x^1)^{-1}
\ee
for the metric (\ref{startmetric}).\\
(ii) After the T-duality, non zero components of the NS-NS 2-field are $B_{23} = - g_{23}/g_{22} = -B_1 $ and $B_{25} = -g_{25}/g_{22} = -B_2$. The NS5-brane charge is given by an integral of the 3-form field strength $H=dB$. 
 We restrict that the NS5-branes charges are measured on the outside of the diamond. 
This is because the `origin' of the deformed conifold is on
 the 3-cycle with non-zero radius. The deformed conifold metric is 
 not defined inside the 3-cycle. 
 The size of 3-cycle corresponds to the size of the diamond. 
Therefore we expect that the NS5-branes with the diamond metric inspired by the
 deformed conifold metric is also defined only on the outside of the diamond. 
Since  we consider the case that all directions except for the one direction are   smeared, we assume that the components of the field strength are constants. So the components of the NS-NS 2-form linearly depend  on the coordinates, 
$
B_{23} = \sum_{i=3}^{6} n_i\, x^i$ and $
B_{25} = \sum_{i=3}^6 m_i\, x^i$. 
We can set $n_3, m_5, m_3=0$ by using a gauge transformation. Thus we have 
\ben
B_1 &=& -(n_4\, x^4 + n_5\, x^5 + n_6\, x^6), \\
B_2 &=& -(m_4\, x^4 + m_6\, x^6).
\een
Note that we still consider  a single diamond, but 
we introduce $n_i$ and $m_i$. This means that some individual NS5-branes 
wrap the same diamond. 

Let us consider the IIA theory compactified on the 6-dimensional metric. 
Since there are no gauge fields and dilaton, we can trivially lift to the 11 dimensional theory. Therefore we equivalently consider unbroken supersymmetry in the 11-dimensional theory for simplicity.

The condition of preserved supersymmetry is that the supersymmetric variation 
 with respect to  the gravitino must vanish, namely $\delta \psi_\mu =0$, 
 in a vanishing gravitino
 background, 
\ben
\delta \psi_\mu &=& D_\mu \eta + \frac{1}{288}(\Gamma_{\mu\nu\rho\sigma\lambda}
-8 g_{\mu\nu} \Gamma_{\rho\sigma\lambda})F^{\nu\rho\sigma\lambda}\eta ,\\
D_\mu \eta &=& \del_\mu \eta + \frac{1}{4}\omega_\mu{}^{\hat{a}\hat{b}}\Gamma_{\hat{a}\hat{b}} \eta ,
\een
where $F$ is a 4-form field strength and $\omega_\mu{}^{\hat{a}\hat{b}}$ is 
a spin connection. The Majorana spinor $\eta $ is the supersymmetry parameter. 
The hatted indices refer to the $D=11$ tangent space, $\eta_{\hat{a}\hat{b}} =$ diag $(-,+,\cdots, +) $, and $\Gamma_{\hat{a}} $ are the $D=11$ Dirac matrices obeying
\be
\{ \Gamma_{\hat{a}}, \Gamma_{\hat{b}}\} = 2 \eta_{\hat{a}\hat{b}},
\ee
and
\be
\Gamma_{\hat{a_1} \ldots \hat{a_n}} = \Gamma_{[\hat{a_1}} \ldots \Gamma_{\hat{a_n}]}\, .
\ee
Since there are no 4-form field strength in our case, the condition reduces to
\be
D_\mu \eta =0\, .
\ee
We choose  normal coordinate basis $\theta^{\hat{a}}$ of the metric in  
11-dimensions  as
\ben
\theta^{\hat{1}} &=& A \, dx^1 ,\\
\theta^{\hat{2}} &=& A^{-1}\, (dx^2 - (n_4\, x^4 + n_5\, x^5 + n_6\, x^6)\, dx^3 -(m_4\, x^4 + m_6\, x^6)\,dx^5) , \\
\theta^{\hat{3}} &=& \sqrt{C-D}\, ( -\sin\al\, dx^3 -\cos\al\, dx^4 + dx^6),\\
\theta^{\hat{4}} &=& \sqrt{C-D}\, (\cos\al\, dx^3 - \sin\al\, dx^4 + dx^5 ),\\
\theta^{\hat{5}} &=& \sqrt{C+D}\, (\sin\al\, dx^3 + \cos\al\, dx^4 + dx^6 ),\\
\theta^{\hat{6}} &=& \sqrt{C+D}\, (-\cos\al\, dx^3 + \sin\al\, dx^4 + dx^5),\\
\theta^{\hat{i}} &=& dx^i\quad i = 0,7,\cdots ,10.
\een
The metric of the six-dimensional curved space (\ref{startmetric})  is given by $\sum_{i=1}^6 (\theta^{\hat{i}})^2$. 
Here we assume that 
$C-D$ and
$C+D$  are positive to make the metric be Lorentzian.

Supposing that the supersymmetry parameter $\eta$ depends only on the $x^1$, the non zero components of $D_\mu \eta $ relative to the metric (\ref{startmetric}) are
\ben
D_1 \eta &=& \del_1 \eta, \label{d1e}\\
D_2 \eta &=&  \frac{2\, f}{A^2}\, \eta , \label{condition2}\\
D_3 \eta &=& \left(- \frac{2\, (n_4 x^4+n_5 x^5+n_6 x^6)}{A^2} f + 
\frac{1}{A} \, g_1 + \frac{1}{A}\, g_2 \right)\,  \eta, \label{condition3}\\
D_4 \eta &=& \left( \frac{1 }{A}\, g_3
+ \frac{1}{A}\, g_4 \right) \, \eta  ,\\
D_5 \eta &=& \left( -\frac{2\,(m_4 x^4+m_6 x^6)}{A^2} f 
+ \frac{1}{A}h_1 \right) \, \eta ,\\
D_6 \eta &=& \frac{1}{A} h_2\, \eta , \label{condition6}
\een
where
\ben
f &=& \frac{A'}{A} \Gh12 + X_1 \Gh34 + X_2 \Gh35 + X_3 \Gh36 
 -X_4\Gh45 - X_5 \Gh46 - X_6 \Gh56, \label{eqf}\\
g_1 &=& Y_1 \Gh31 + Y_2 \Gh51 + Z_1 \Gh42 + Z_2 \Gh62 , \\
g_2 &=& Y_3 \Gh41 + Y_4 \Gh61 + Z_3 \Gh32 + Z_4 \Gh52 , \\
g_3 &=& -Y_3 \Gh31- Y_4 \Gh51 + Z_5 \Gh42 + Z_6 \Gh62 , \\
g_4 &=& Y_1 \Gh41+ Y_2 \Gh61 + Z_7 \Gh32 + Z_8 \Gh52, \\
h_1 &=& V_1 \Gh41 + V_2 \Gh61 + W_1 \Gh32 + W_2 \Gh42 + W_3 \Gh52 + W_4 \Gh62 ,\\
h_2 &=& V_1 \Gh31 + V_2 \Gh51 + W_5 \Gh32 + W_6 \Gh42 + W_7 \Gh52 + W_8 \Gh62 ,
\een
and $X,Y,Z,V,W$'s are defined as
\ben
&&X_1 = \frac{m_6-n_4 + (n_6-m_4) \cos\al + n_5 \sin\al}{8(C-D)}, \quad
X_2 = \frac{n_6 \sin\al}{4\sqrt{C^2-D^2}}, \non \\
&&X_3 = \frac{m_6 + n_4-(m_4+n_6)\cos\al + n_5\sin\al}{8\sqrt{C^2-D^2}}, \non \\
&&X_4 = \frac{m_6 + n_4+ (m_4+n_6)\cos\al - n_5\sin\al}{8\sqrt{C^2-D^2}},\non \\
&&X_5 = \frac{n_5 \cos\al + m_4\sin\al}{4\sqrt{C^2 -D^2}},\quad 
X_6 = \frac{-m_6 + n_4 + (n_6-m_4)\cos\al + n_5\sin\al}{8(C+D)},\non \\
&&Y_1 = -\frac{(C'-D')\sin\al}{\sqrt{C-D}},\qquad
Y_2 = \frac{(C'+D')\sin\al}{\sqrt{C+D}}, \non \\
&&Y_3 = \frac{(C'-D')\cos\al}{\sqrt{C-D}},\qquad
Y_4 = -\frac{(C'+D')\cos\al}{\sqrt{C+D}}, \non \\
&&Z_1 = \frac{n_5-n_4\sin\al}{2\sqrt{C-D}},\qquad
Z_2 = \frac{n_5 + n_4\sin\al}{2\sqrt{C+D}},\non \\
&&Z_3 = \frac{n_6-n_4\cos\al}{2\sqrt{C-D}},\qquad
Z_4 = \frac{n_6 + n_4\cos\al}{2\sqrt{C+D}},\non \\
&&Z_5 = -\frac{m_4 +n_4 \cos\al}{2\sqrt{C-D}},\qquad
Z_6 = -\frac{m_4 -n_4 \cos\al}{2\sqrt{C+D}},\non \\
&&Z_7 = \frac{n_4 \sin\al}{2\sqrt{C-D}},\qquad
Z_8 = -\frac{n_4\sin\al}{2\sqrt{C+D}},\non \\
&&V_1 = \frac{C'-D'}{\sqrt{C-D}},\qquad
V_2 = \frac{C'+D'}{\sqrt{C+D}},\non \\
&&W_1 = \frac{m_6-m_4\cos\al+n_5\sin\al}{2\sqrt{C-D}},\qquad
W_2 = -\frac{n_5\cos\al + m_4 \sin\al}{2\sqrt{C-D}},\non \\
&&W_3 = \frac{m_6 + m_4 \cos\al - n_5\sin\al}{2\sqrt{C+D}},\qquad
W_4 = \frac{n_5\cos\al + m_4\sin\al}{2\sqrt{C+D}},\non \\
&&W_5 = \frac{n_6\sin\al}{2\sqrt{C-D}},\qquad
W_6 = -\frac{m_6 + n_6 \cos\al}{2\sqrt{C-D}},\non \\
&&W_7 =-\frac{n_6 \sin\al}{2\sqrt{C+D}},\qquad
W_8 = -\frac{m_6-n_6\cos\al}{2\sqrt{C+D}}. \non 
\een

From eq.(\ref{d1e}), $\eta $ must be a constant spinor. 
When $D=0,\, n_5=n_6=m_4=0 ,\, n \equiv n_4=m_6$ and $A = 2 C = 1 + n\, |x^1|$, the metric (\ref{startmetric}) becomes the T-dual of the usual intersecting 
$n$ NS5-NS5' branes solution. 
  Indeed  $1/4$ supersymmetries are preserved. For such solution, for example, eq.(\ref{eqf}) becomes
\ben
f &=& \frac{n}{2 A} \left( 2 \Gamma_{\hat{1}\hat{2}}
-  \Gamma_{\hat{4}\hat{5}} +  \Gamma_{\hat{3}\hat{6}} \right) \non \\
&=& \frac{n}{2 A}\Gamma_{\hat{1}\hat{2}} \left( (1 + \Gamma_{\hat{1}\hat{2}\hat{4}\hat{5}}) + 
(1 - \Gamma_{\hat{1}\hat{2}\hat{3}\hat{6}}) \right).
\een
So the solution of the condition (\ref{condition2}) is given by 
\be
\eta = \frac{1 - \Gamma_{\hat{1}\hat{2}\hat{4}\hat{5}}}{2}
\frac{1 + \Gamma_{\hat{1}\hat{2}\hat{3}\hat{6}}}{2} \, \tilde{\eta} \label{psusy}
\ee
where $\tilde{\eta}$ is a constant spinor. It is easy to check that the 
solution $\eta $ also satisfy all other conditions 
(\ref{condition3})-(\ref{condition6}). So the solution (\ref{psusy}) has 
unbroken $1/4$ supersymmetries for the intersecting NS5-branes.

The NS5-branes with the diamond must preserve the same supersymmetries as eq.(\ref{psusy}) of intersecting NS5-branes since they are described by the same holomorphic coordinates \cite{BBS}. The conditions for preserving 
supersymmetry are
\ben
&&\frac{A'}{A} = 2\, X_3 = 2\, X_4, \\
&&X_1 = X_2 = X_5 = X_6 =0,\\
&&Y_1 = - Z_2=Z_8, \qquad Y_2 = -Z_1=Z_7,\\
&&Y_3 = Z_4 = Z_6,\qquad Y_4 = Z_3=Z_5, \\
&& V_1 = W_3=-W_8, \qquad V_2 = W_1=-W_6, \\
&& W_2 = W_4 = W_5 =W_7 =0.
\een
First,  we find $n_4 = m_6$ from $X_1 = X_6=0$, then define 
$n\equiv n_4 = m_6$. 
Those equations are solved for
\ben
A = 2\,C &=& H , \\
2\, D &=& \beta \, H ,\\
n_5 = n_6 = m_4 &=&0, 
\een
where
\be
H \equiv 1 + \frac{n}{\sqrt{1-\beta^2}}\, |x^1|.
\ee
Here $\beta $ is one of integral constants and others are fixed to determine 
asymptotic behavior of $H$. From the condition that both $C+D$ and $C-D$ are 
positive, the range of $\beta $ is restricted within 
\be
|\beta | < 1.
\ee
When $\beta$ vanishes, the solution smoothly reduce to the T-dual of intersecting NS5-branes.

Thus we found the solution preserving $1/4$ supersymmetries.  The metric for 
the compactified space is
\be
ds^2 = H^2\,  (dx^1)^2 + H^{-2}\,  ds_1^2 + H \, ( ds_2^2 + \beta\, ds_3^2)  \label{metric}
\ee
where
\ben
\hspace{-0.8cm}ds_1^2 &=& (dx^2 - n\, x^4\, dx^3  -n\, x^6\, dx^5)^2 ,\\
\hspace{-0.8cm}ds_2^2 &=& \sum_{i=3}^6 (dx^i)^2,\\
\hspace{-0.8cm}ds_3^2 &=& 2( \sin\al\, (dx^3\,dx^6 + dx^4\, dx^5) + \cos\al (dx^4\, dx^6 - dx^3\, dx^5 )), \\
\hspace{-0.8cm}H &=& 1 + \frac{n}{\sqrt{1-\beta^2}} \, |x^1|, \quad |\beta | < 1 .
\een
The metric (\ref{metric}) is a non-compact CY metric.
It is straightforward to check that the metric  is indeed Ricci flat.

The determinant of the metric (\ref{metric}) is proportional to
$(1-\beta^2)^2$ which
 vanishes if $\beta^2 \to 1$. It seems to be the same as the deformation of conifold case where the determinant of the metric vanishes on the $S^3$ surface. However, since the metric (\ref{metric}) is  smeared, the  meaning of this limit  is not so clear. We will discuss the meanings soon later.

Let us consider the compactification of the 10-dimensional type II theory
on the 6-dimensional space (\ref{metric})
\footnote{
We may consider both the IIA and IIB theory, since there are no R-R fields.}, 
and take the T-duality along the $x^2$-direction. The duality relations are
 proposed in \cite{BHO}. We have the following components relative to the metric (\ref{metric}):
\ben
j_{mn} &=& g_{mn}-\frac{g_{2m}\, g_{2n}}{g_{22}}, \quad 
j_{22} = \frac{1}{g_{22}}, \\
B_{2m} &=& -\frac{g_{2m}}{g_{22}} , \quad
e^{2\phi} = \frac{1}{g_{22}}. 
\een
Here $m,n$ are $0,1,3,\cdots , 9$, and $\phi$ is the dilation of the theory 
after the T-duality.  The metric becomes
\be
ds^2 = -(dx^0)^2 + H^2\, \sum_{i=1}^2 (dx^i)^2 + H \, ( \sum_{i=3}^6 (dx^i)^2  + \beta\, ds_3^2) + 
\sum_{i=7}^9 (dx^i)^2 \label{diametric}
\ee
where
\be
ds_3^2 = 2( \sin\al\, (dx^3\,dx^6 + dx^4\, dx^5) + \cos\al (dx^4\, dx^6 - dx^3\, dx^5 )), \label{offdiag}\\
\ee
and the dilaton is given by
\be
e^{2\, \phi} = H^2.
\ee
The non-vanishing components of the NS-NS 3-form field strength become
\be
H_{234} = H_{256} = n. \label{strength}
\ee
The metric (\ref{diametric}) reduces to the intersecting NS5-NS5'-branes metric smoothly if $\beta = 0$, and the preserved supersymmetries are the same as intersecting ones. Therefore we found the metric of the NS5-branes with a diamond smeared except for the $x^1$-direction  and the parameter $\beta $ corresponds to the size of the diamond
\footnote{ More precisely  $\sqrt{|\beta|}$ can be
identified with the size of the diamond, as we 
will see in sec \ref{secbx}.}.

Let us consider  the large size of the diamond 
 $\beta^2 \to 1$. Then the determinant of the metric vanishes and the metric 
is singular.  This is because we have made the metric with the ansatz
 that the charges of NS5-branes are measured outside  the diamond. The ansatz
  is broken if  the point where the charges are measured meets with 
  the diamond.
  This is a nice correspondence with
  the deformed conifold metric  
whose  determinant vanishes if one goes on the $S^3$ surface, $\tau \to 0$ as we have seen in sec. \ref{secdefconi}.

The metric (\ref{diametric}) describes the NS5-branes with the diamond 
which is in the delocalized directions. 
Let us see that the metric (\ref{diametric}) can transform to the 
ordinary intersecting one and the diamond shrinks.

First, the parameter $\al $ is an angle of  a coordinate rotation on either 
the $(x^3,x^4)$ or $(x^5,x^6)$  plane. 
If  we choose the $(x^3, x^4)$ plane
\footnote{
Even if we choose the $(x^5,x^6)$ plane, the following discussion is
appropriate to the plane.
}, we can remove it by the following 
 rotation of the coordinates:
\ben
x^3 &\to & \cos\al\, x^3 + \sin\al\, x^4 \\
x^4 &\to &-\sin\al\, x^3 + \cos\al\, x^4.
\een
The 3-form field strength (\ref{strength}) does not change by the rotation. 
The off-diagonal part in the metric (\ref{offdiag}) becomes 
\be
ds_3^2 \to 2\, (dx^4\, dx^6 - dx^3 \, dx^5),
\ee
and other parts do not change. 
Thus the parameter $\al $ is removed from the solution of NS5-branes with the diamond.

The metric can be  diagonalized by the following coordinate transformation:
\ben
x^3 &\to & x^3 + \beta\, x^5, \label{tf1}\\
x^6 &\to & x^6 - \beta\, x^4. \label{tf2}
\een
The terms which  associates with the $(x^3,x^4,x^5,x^6)$ directions in 
the metric become
\be
ds_{3456}^2 \equiv H\, \left( (dx^3)^2 + (1-\beta^2) \left((dx^4)^2 +  (dx^5)^2
\right) + (dx^6)^2 \right).
\ee
The components of field strength do not change by the transformation 
(\ref{tf1}) and (\ref{tf2}). Finally we rescale the $(x^4,x^5)$ coordinates as
\be
\ba{l}
x^4 \to  \frac{x^4}{\sqrt{1-\beta^2}}, \\
x^5 \to  \frac{x^5}{\sqrt{1-\beta^2}},
\ea \label{rescale1}
\ee
and  the parameter $n$ as 
\be
n \to  \sqrt{1-\beta^2}\, n. \label{rescale2}
\ee
Then we have the intersecting NS5 solution
\be
ds^2 = -(dx^0)^2 + H^2\, \sum_{i=1}^2 (dx^i)^2+ 
H \,  \sum_{i=3}^6 (dx^i)^2 + \sum_{i=7}^9 (dx^i)^2 \label{intersect}
\ee 
where
\be
H= 1 + n\, |x^1|.
\ee
The components of 3-form field strength are the same  as (\ref{strength})
 due to the cancelation of the rescale (\ref{rescale1}) and (\ref{rescale2}). 
Thus the metric (\ref{diametric}) reduce to the intersecting one and 
 the diamond shrinks.

We can  see the effect of the transformation by using the duality relation 
\cite{BHO}. Since  we apply it along the transformed coordinates, 
the transformation 
correspond to changing the direction of the compactification.

\section{U-dualities \label{secdual}}

The (deformed) conifold geometry or NS5-branes configurations are related to 
other interesting configurations by dualities. In this section we consider the
consistency under the dualities. 

\subsection{NS5-brane and D4-brane \label{nsd}}

We consider the type IIB theory on the conifold. The T-duality along the $x^2$ direction  take us the configuration of intersecting NS5-NS5' branes  in the type IIA theory whose world-volume directions are
\be
\ba{c|cccccccccc}
NS5& (0&&&3&4&&&7&8&9) \\
NS5'& (0&&&&&5&6&7&8&9 ).
\ea \label{start}
\ee

We first lift the configuration to the M-theory and
 flip the $x^5$ and $x^{10}$ directions.

\be
\ba{c|ccccccccccc}
M5& (0&&&3&4&&&7&8&9&) \\
M5'& (0&&&&&&6&7&8&9&10 ).
\ea 
\ee
We dimensionally reduce the $x^{10}$ direction and  have the intersecting NS5 and D4-branes
 whose world-volume directions are
\be
\ba{c|cccccccccc}
NS5& (0&&&3&4&&&7&8&9) \\
D4 & (0&&&&&&6&7&8&9 )
\ea \label{step1}
\ee
in the type IIA theory since one of M5-branes wraps on the compactified circle.

The D4-brane ends on the NS5-brane and
each open D4-branes can slide along the NS5-brane \cite{strominger}. 
The shift of 
 the open D4-branes in the $(x^3,x^4)$ space corresponds to the size of the diamond. This can be seen as follows. 
  We introduce the holomorphic coordinate $x = x^3 + i\, x^4 $. Let the 
  position of the one of the open D4-branes on the $x$ plane be $x=0$ and 
  the other be $x=m$.
 In M-theory, these branes are described by the M5-brane wrapping on the 
 holomorphic curve
 \cite{witten}
\be
x\, t - (x-m) =0, \label{m5}
\ee
where $t=e^y$, $y=x^6+ ix^{10}$ and we take the radius of 
the $x^{10}$-direction  as 1.   
We choose  coefficients of the curve (\ref{m5}) as that 
the solutions of the curve are $x=0, y=0$ when $m=0$.
 In the near end points of D4-branes limit $|y| <\! < 1$, 
 since $t \sim 1 + y$,  
 the curve (\ref{m5}) becomes the same as the curve (\ref{dia}). Therefore we 
 find that  the squareroot of $|m|$ corresponds to  the size of the 
 diamond $\epsilon$.

This correspondence is  generalized to intersecting 
$n$ NS5-branes and $m$ NS5'-branes in IIA theory \cite{AKLM}. 
In this case, we have a generalized conifold $({\cal G}_{mn}:uv=x^m y^n)$ after the T-duality along the overall transverse direction to the NS5 and NS5'-branes.
The generalized conifold can be deformed to a smooth space by 
\be
uv = \sum_{i,j=0}^{n,m} m_{ij} x^i\, y^j. \label{geneconi}
\ee
So,  after the T-duality,  we have the curve 
\ben
0 &=& \sum_{i,j=0}^{n,m} m_{ij} x^i\, y^j \non \\
  &=& (m_{m\, n}x^m + \cdots m_{0\, n})\, y^n 
  + (m_{m\, n-1}x^m + \cdots m_{0\, n-1})\, y^{n-1}+
 \cdots \non \\
 &&\quad +  (m_{m\, 0}x^m + \cdots m_{0\, 0})   \label{mij}
\een
on which a single NS5-brane wraps. 
The parameters $m_{ij}$ correspond to the locations of the original NS5 and
 NS5'-branes and the size of diamonds opened at each intersecting points. 
If we lift to the M-theory, we have a single M5-brane wrapping the 
Seiberg-Witten curve which is the same as the curve (\ref{mij}). 
We have an $N=2$ four-dimensional $SU(m)^{n-1}$ 
gauge theory with  vanishing beta functions on D4-branes  after the coordinate 
flip and the dimensional reduction along the 
direction in originally  the NS5'-branes. 
 The matters consist of the  $(n-2)$ 
hypermultiplets in the bi-fundamental representation 
and two hypermultiplets in 
  the fundamental representation of $SU(m)$, which come from 
 the semi-infinite D4-branes in the left and right. 
Thus the size of diamonds or locations of NS5 and NS5'-branes
 corresponds to the moduli parameters of the $N=2$ gauge theory. 
In fact the $(m-1) + (n-1)$  relative branes positions and 
the size of diamonds at $mn$ intersection are mapped to 
$(n-2)+2m$ bare masses of hypermultiplets, $(n-1)(m-1)$ vevs of scalars 
in adjoint representation of the gauge groups $SU(m)^{n-1}$ and $n-1$ 
complex gauge 
coupling constants. Since all beta-functions are zero, 
gauge coupling constants 
are also moduli parameters. 
As a result, we have $mn+ m+ n-2$ parameters in total. This number exactly 
agrees with the deformation parameter in eq.(\ref{geneconi}).

\subsection{Other U-dualities \label{secbx}}

Let us consider the correspondence between the displacement of the 
ends of D4-branes 
and the size of the diamond by using the NS5-brane with the diamond metric (\ref{diametric}) .

We consider the following duality maps. First, we start with the configuration 
of intersecting NS5 and D4-branes,
\be
\ba{c|cccccccccccc}
NS5 &(&0&&&3&4&&&7&8&9&) \\
D4 &(&0&&&&&&6&7&8&9&). \label{step2-1}
\ea
\ee
Secondly, the T-duality along the $x^7,x^8$ and $x^9$-directions maps
the configuration to
\be
\ba{c|cccccccccccc}
NS5 &(&0&&&3&4&&&7&8&9&) \\
D1 &(&0&&&&&&6&&&&).
\ea \label{step2}
\ee
Finally, we apply S-dual operation and  and obtain
\be
\ba{c|cccccccccccc}
D5 &(&0&&&3&4&&&7&8&9&) \\
F1 &(&0&&&&&&6&&&&). \label{last}
\ea
\ee

It is shown in \cite{DHL} that if there is the displacement of the 
two ends of the fundamental string along
the $x^i$-direction on the D-brane, the displacement becomes a constant 
$B$-field after the T-duality along the direction. 
The relation of 
the displacement $\delta x^i$ and the component of the $B$-field is given by
\be
\delta x^i = B_{i6}
\ee
where the fundamental string extends along  the $x^6$-direction. 
Therefore if we take the magnitude of the displacement along the $x^3,x^4$-directions as $\delta x^3, \delta x^4$ respectively, we have components of the $B$-field such that
\be
\ba{l}
B_{36} = \delta x^3 \\
B_{46} = \delta x^4
\ea \label{bandx}
\ee
after the T-duality along the $x^3$ and $x^4$-directions.

It is apparent that the displacement of the end points of D4-branes
 on the $(x^3,x^4)$ directions is the same as that of fundamental strings 
 in the directions under the duality map from the 
 configuration (\ref{step2-1}) 
 to (\ref{last}). 
The diamond size corresponds to the 
$B$-field since the displacement $\delta x=m$ corresponds to it as we 
discussed in the previous subsection. We  confirm the observation by using the NS5-branes with a diamond metric (\ref{diametric})  and the duality relations given in \cite{BHO}. We can also identify with the relation between the parameter $\beta $ 
and the size of the diamond. 
 
We trace the duality chain (\ref{start})-(\ref{step1}),
(\ref{step2}) and (\ref{last}) and take the T-dualities to the configuration 
(\ref{last}) along the $x^3,x^4$ direction. 
Applying the dualities relations to the metric (\ref{diametric}), we obtain
\ben
B_{36} &=& \beta\, \cos\al \\
B_{46} &=&-\beta\, \sin\al.
\een

From (\ref{bandx}) we have
\be
\delta x\; (=m) = \beta\, \cos\al - i\, \beta\, \sin\al.
\ee
Therefore we find that $\sqrt{|m|} = \sqrt{|\beta|}$ is the size of the diamond. The  direction of 
the displacement is rotated in  the $(x^3,x^4)$ directions at 
the angle $\al $. 
 We confirm 
 the discussion in the sec \ref{secmetric}.


\section{Conclusion and Discussion \label{secconclusion}}
We start with the metric  inspired by the
deformed conifold metric, and
 obtain the NS5-branes with the diamond metric which
 is smeared except for one of the overall transverse directions. 
The metric is also obtained  from the intersecting 
NS5-branes one 
by the transformation (\ref{tf1}) and (\ref{tf2}). It means that
  the diamond is spread by the transformation. 
The parameter $\beta $ in the metric is the square of the size of the diamond 
and $\al $ is the rotation angle on the plane over which originally one of 
intersecting NS5-branes extends. 
The NS5-brane with the diamond relates to the NS5-brane and the D4-brane via 
string dualities, where the D4-brane breaks on the NS5-brane. The 
displacement of the end points  is equal to $\beta $.

The metric of NS5-branes with the diamond is delocalized and does not have 
local information 
about the diamond. However we can get the information about the size 
of the diamond because we assume that there is only one diamond. If there 
are more than one diamond as in sec. \ref{nsd}, 
 it is difficult to obtain  information
 about each diamonds from smeared metrics.

 The  partially  localized  
 solutions for the intersecting branes are 
 obtained \cite{youm, FS, loewy, MP2}. 
 The fully localized solution for a  M5-brane wrapping a Riemann surface 
 is also discussed in \cite{FS}. The fully localized 
 solution is presented with a  K\"ahler potential, 
 however, the explicit form of the K\"ahler potential is not 
 known.  In \cite{GKMT}, the authors discuss the K\"ahler potential
 perturbatively for various types of intersecting branes and  branes wrapping 
  curves,  
 and find that the perturbation theory is well 
 behaved  at least to the second order if 
 there are less than three overall 
 transverse dimensions.

   It is an interesting  problem how the localized intersecting branes
  solutions can be deformed  and how the conifold or 
  the deformed  conifold relate to such solutions   
  by the duality map.

\section*{Acknowledgments}
The authors would like to thank  Toshio Nakatsu for valuable discussions. 
T.Y also acknowledges  Koichi Murakami and Toshihiro Matsuo for very useful
 discussions.
 T.Y is supported in part by the JSPS Research Fellowships.


\begin{thebibliography}{99}



\bibitem{AK}
A. Kehagias, 
``New Type IIB Vacua and their F-Theory Interpretation'',
\PL{B435}{98}{337},
hep-th/9805131.

\bibitem{KW}
I. R. Klebanov and E. Witten, 
``Superconformal Field Theory on Threebranes at a Calabi-Yau Singularity'',
\NP{B536}{98}{199},
hep-th/9807080.


\bibitem{AFHS}
BS. Acharya, JM. F. O'Farrill, CM. Hull and B. Spence, 
``Branes at conical singularities and holography'',
{\rm Adv.Theor.Math.Phys.} {\bf 2} (1999) 1249,
hep-th/9808014.

\bibitem{MP}
D. R. Morrison and M. R. Plesser, 
``Non-Spherical Horizons, I'',
{\rm Adv.Theor.Math.Phys.} {\bf 3} (1999) 1-81,
hep-th/9810201.

\bibitem{maldacena}
J.M. Maldacena, 
``The Large N Limit of Superconformal Field Theories and Supergravity'',
{\rm Adv.Theor.Math.Phys.} {\bf 2} (1998) 231, {\rm Int.J.Theor.Phys.}{\bf 38} (1999) 1113,
hep-th/9711200.

\bibitem{AGMOO}
O. Aharony, S.S. Gubser, J. Maldacena, H. Ooguri and Y. Oz, 
``Large N Field Theories, String Theory and Gravity'',
hep-th/9905111.

\bibitem{DM}
K. Dasgupta and S. Mukhi, 
``Brane Constructions, Conifolds and M-Theory'',
\NP{B551}{99}{204},
hep-th/9811139.

\bibitem{Uranga}
A. M. Uranga, 
``Brane Configurations for Branes at Conifolds'',
\JHEP{9901}{99}{022},
hep-th/9811004.

\bibitem{witten}
E. Witten, 
``Solutions Of Four-Dimensional Field Theories Via M Theory'',
\NP{B500}{97}{3},
hep-th/9703166.

\bibitem{Unge}
R. von Unge, 
``Branes at Generalized Conifolds and Toric Geometry'',
\JHEP{9902}{99}{023},
hep-th/9901091.

\bibitem{AKLM}
M. Aganagic, A. Karch, D. Lust, A. Miemiec, 
``Mirror Symmetries for Brane Configurations and Branes at Singularities'',
hep-th/9903093.

\bibitem{TO}
K. Oh and R. Tatar, 
``Branes at Orbifolded Conifold Singularities and Supersymmetric Gauge Field Theories'',
\JHEP{9910}{99}{031},
hep-th/9906012.

\bibitem{BSV}
M. Bershadsky, V. Sadov and C. Vafa, 
``D-Strings on D-Manifolds'',
\NP{B463}{96}{398},
hep-th/9510225.


\bibitem{HZ}
A. Hanany andA. Zaffaroni, 
``On the realization of chiral four-dimensional gauge theories using branes'',
\JHEP{9805}{98}{01},
hep-th/9801134.

\bibitem{HU}
A. Hanany and A. M. Uranga, 
``Brane Boxes and Branes on Singularities'',
\JHEP{9805}{98}{013},
hep-th/9805139.


\bibitem{STZ}
A. Strominger, S. T. Yau and E. Zaslow, 
``Mirror Symmetry is T-Duality'',
\NP{B479}{96}{243},
hep-th/9606040.

\bibitem{CO}
P. Candelas and C. de la Ossa, 
``Comments on conifolds'',
\NP{B342}{90}{246}.


\bibitem{MT}
 R. Minasian and D. Tsimpis, 
 ``On the geometry of non-trivially embedded branes'',
hep-th/9911042.

\bibitem{KLMVW}
A. Klemm, W. Lerche, P. Mayr, C.Vafa and N. Warner, 
``Self-Dual Strings and N=2 Supersymmetric Field Theory'',
\NP{B477}{96}{746},
hep-th/9604034,\\
S. Katz, P. Mayr and  C. Vafa, 
``Mirror symmetry and Exact Solution of 4D N=2 Gauge Theories I'',
\ATMP{1}{98}{53},
hep-th/9706110. 

\bibitem{BHO}
E. Bergshoeff, C.M. Hull and T. Ortin, 
``Duality in the Type--II Superstring Effective Action'', \NP{B451}{95}{547},
hep-th/9504081.

\bibitem{BBS}
K. Becker, M. Becker and A. Strominger, 
``Fivebranes, Membranes and Non-Perturbative String Theory'',
\NP{B456}{95}{130},
hep-th/9507158.

\bibitem{strominger}
A. Strominger, 
``Open P-Branes'',
\PL{B383}{96}{44},
hep-th/9512059.

\bibitem{DHL}
M. R. Douglas, C. Hull, 
``D-branes and the Noncommutative Torus'',
\JHEP{9802}{98}{008},
hep-th/9711165.
\\
M. Li, 
``Comments on Supersymmetric Yang-Mills Theory on a Noncommutative Torus'',
hep-th/9802052.

\bibitem{youm}
D. Youm, 
``Partially Localized Intersecting BPS Branes'',
hep-th/9902208.


\bibitem{FS}
A. Fayyazuddin and D. J. Smith, 
``Localized intersections of M5-branes and four-dimensional superconformal field theories'',
\JHEP{9904}{99}{030},
hep-th/9902210.

\bibitem{loewy}
A. Loewy, 
``Semi Localized Brane Intersections in SUGRA'',
\PL{B463}{99}{41},
hep-th/9903038.

\bibitem{MP2}
D. Marolf and A. W. Peet, 
``Brane Baldness vs. Superselection Sectors'',
\PR{D60}{99}{105007},
hep-th/9903038.

\bibitem{GKMT}
A. Gomberoff, D. Kastor, D. Marolf and J. Traschen, 
``Fully Localized Brane Intersections - The Plot Thickens'',
{\rm Phys.Rev.} {\bf D61} (2000) 024012,
hep-th/9905094.

\end{thebibliography}
\end{document}